\documentclass[aps,prd,twocolumn,superscriptaddress,showpacs,showkeys,
nofootinbib]{revtex4-1}

\usepackage{latexsym}
\usepackage{amsmath}
\usepackage{amssymb}
\usepackage{amsfonts}

\usepackage{color}

\usepackage{supertabular} 
\usepackage{placeins}
\usepackage{epsfig}
\usepackage{graphicx}

\begin{document}

%\preprint{}

\title{Molecular components in P-wave charmed-strange mesons}


\author{Pablo G. Ortega}
%\thanks{}
\altaffiliation{Currently at Instituto de F\'isica Corpuscular (IFIC), 
CSIC-Universidad de Valencia, E-46071 Valencia, Spain}
\affiliation{CERN (European Organization for Nuclear Research), CH-1211 Geneva, 
Switzerland}

\author{Jorge Segovia}
%\thanks{}
\affiliation{Physik-Department, Technische Universit\"at M\"unchen, 
James-Franck-Str. 1, 85748 Garching, Germany}

\author{David R. Entem}
%\thanks{}
%\affiliation{}

\author{Francisco Fern\'andez}
%\thanks{}
\affiliation{Grupo de F\'isica Nuclear and Instituto Universitario de F\'isica 
Fundamental y Matem\'aticas (IUFFyM), Universidad de Salamanca, E-37008 
Salamanca, Spain}


\date{\today}

\begin{abstract}
Results obtained by various experiments show that the $D_{s0}^{\ast}(2317)$ and 
$D_{s1}(2460)$ mesons are very narrow states located below the $DK$ and 
$D^{\ast}K$ thresholds, respectively. This is markedly in contrast with the 
expectations of naive quark models and heavy quark symmetry. Motivated by a 
recent lattice study which addresses the mass shifts of the $c\bar{s}$ ground 
states with quantum numbers $J^{P}=0^{+}$ ($D_{s0}^{\ast}(2317)$) and 
$J^{P}=1^{+}$ ($D_{s1}(2460)$) due to their coupling with $S$-wave 
$D^{(\ast)}K$ thresholds, we perform a similar analysis within a 
nonrelativistic constituent quark model in which quark-antiquark and 
meson-meson degrees of freedom are incorporated. The quark model has been 
applied to a wide range of hadronic observables and thus the model parameters 
are completely constrained. The coupling between quark-antiquark and 
meson-meson Fock components is done using a $^{3}P_{0}$ model in which its only 
free parameter $\gamma$ has been elucidated performing a global fit to the 
decay widths of mesons that belong to different quark sectors, from light to 
heavy. We observe that the coupling of the $0^{+}$ $(1^{+})$ meson sector to 
the $DK$ $(D^{\ast}K)$ threshold is the key feature to simultaneously lower the 
masses of the corresponding $D_{s0}^{\ast}(2317)$ and $D_{s1}(2460)$ states 
predicted by the naive quark model and describe the $D_{s1}(2536)$ meson as the 
$1^{+}$ state of the $j_{q}^{P}=3/2^{+}$ doublet predicted by heavy quark 
symmetry, reproducing its strong decay properties. Our calculation allows to 
introduce the coupling with the $D$-wave $D^{\ast}K$ channel and the 
computation of the probabilities associated with the different Fock components 
of the physical state.
\end{abstract}

\pacs{12.39.Pn, 14.40.Lb, 14.40.Rt}

\keywords{Potential models, Charmed mesons, Exotic mesons}

\maketitle

%%%%%%%%%%%%%%%%%%%%%%%%%%%%%%%%%%%%%%%%%%%%%%%%%%%%%%%%%%%%%%%%%%%%%%%%%%%%%%%
%%%%%%%%%%%%%%%%%%%%%%%%%%%%%%%%%%%%%%%%%%%%%%%%%%%%%%%%%%%%%%%%%%%%%%%%%%%%%%%


\section{INTRODUCTION}
\label{sec:introduction}

The discovery in $2003$ of the resonances $D_{s0}^{\ast}(2317)$ 
$(J^{P}=0^{+})$~\cite{Aubert:2003fg} and $D_{s1}(2460)$ 
$(1^{+})$~\cite{Besson:2003cp} yielded great interest among theorists and 
experimentalists due to their unexpected low masses and narrow widths. For 
instance, calculations based on quark models~\cite{Godfrey:1986wj, 
Zeng:1994vj, Gupta:1994mw, Ebert:1997nk, Lahde:1999ih, DiPierro:2001dwf} and 
early lattices gauge theories~\cite{Boyle:1997aq, Boyle:1997rk, Hein:2000qu, 
Lewis:2000sv, Bali:2003jv, diPierro:2003iw, Dougall:2003hv} predicted for these 
states masses which were around $100\,{\rm MeV}$ above their respective $DK$ 
and $D^{\ast}K$ thresholds, whereas the experimental results lie $40\,{\rm 
MeV}$ below such thresholds.

Prior to the discovery of these two states, the heavy-light meson sectors were 
reasonably well understood in the $m_{Q}\to\infty$ limit. In such a limit, 
heavy quark symmetry (HQS) holds~\cite{Isgur:1991wq}. The heavy quark acts as a 
static color source, its spin $s_{Q}$ is decoupled from the total angular 
momentum of the light quark $j_{q}$ and they are separately conserved. Then, 
the heavy-light mesons can be organized in doublets, each one corresponding to 
a particular value of $j_{q}$ and parity. For the lowest $P$-wave charmed 
mesons, HQS predicts two doublets which are labeled by $j_q^P=1/2^+$ with 
$J^P=0^{+},\,1^{+}$ and $j_q^P=3/2^+$ with $J^P=1^+,\,2^+$. The members of each 
doublet differ on the orientation of $s_{Q}$ with respect to $j_{q}$ and, in 
the heavy quark limit, are degenerated. Mass degeneracy is broken at order 
$1/m_{Q}$. Moreover, the strong decays of the $D_{(s)J}\,(j_q=3/2)$ proceed 
only through $D$-waves, while the $D_{(s)J}\,(j_q=1/2)$ decays happen only 
through $S$-waves~\cite{Isgur:1991wq}. The $D$-wave decay is suppressed by the 
barrier factor which behaves as $q^{2L+1}$ where $q$ is the relative momentum 
of the two decaying mesons. Therefore, states decaying through $D$-waves are 
expected to be narrower than those decaying via $S$-waves.

The $D_{s0}^{\ast}(2317)$ and $D_{s1}(2460)$ mesons are considered to be the 
members of the $j_q^{P}=1/2^{+}$ doublet and thus being almost degenerated and 
broad. However, neither experimental values of their masses nor their empirical 
widths accommodate into the theoretical expectations.

These results led to many theoretical speculations about the nature of these 
resonances ranging from conventional $c\bar{s}$ 
states~\cite{Fayyazuddin:2003aa, Sadzikowski:2003jy, Lakhina:2006fy} to 
molecular or compact tetraquark interpretations~\cite{Barnes:2003dj, 
Lipkin:2003zk, Bicudo:2004dx, Szczepaniak:2003vy, Browder:2003fk, 
Nussinov:2003uj, Dmitrasinovic:2005gc}. More recently, a chiral unitary theory 
in coupled channels explains that the $D_{s0}^{\ast}$ state is produced 
dynamically by means of the coupled channels $DK$ and 
$D_{s}\eta$~\cite{MartinezTorres:2011pr, Gamermann:2006nm}. The analysis of the 
$D_{s0}^{\ast}(2317)$ meson's properties using other dynamical coupled-channel 
approaches for meson-meson in $S$-wave can be found in 
Refs.~\cite{Doring:2011ip, Liu:2012zya, Guo:2015dha}. In 
Ref.~\cite{Gamermann:2007fi}, a coupled-channel calculation of 
pseudo-scalar--vector mesons has been performed in order to study the 
$D_{s1}(2460)$ and $D_{s1}(2536)$ states. They found masses $2455\,{\rm MeV}$ 
and $(2573.62-i0.07)\,{\rm MeV}$, respectively. The second state couples mainly 
to $DK^{\ast}$, however the width is very small due to the fact that 
$D^{\ast}K$ in $D$-wave is not included. A molecular interpretation of the 
$D_{s2}^{\ast}(2573)$ has been given in Ref.~\cite{Molina:2010tx}.

Certainly quark models predict $c\bar s$ ground states with quantum numbers 
$J^{P}=0^{+}$ and $1^{+}$ that do not fit the experimental data. As the 
predictions of the quark models are roughly reasonable for other states in the 
charmed-strange sector~\cite{Segovia:2015dia, Song:2015nia}, one must expect 
that the $D_{s0}^{\ast}(2317)$ and $D_{s1}(2460)$ resonances should be 
modifications of the genuine $c\bar s$ states rather than new states out of the 
systematics of the quark model. On this respect, particularly relevant was the 
suggestion~\cite{vanBeveren:2003kd, vanBeveren:2003jv} that the 
coupling of the $J^{P}=0^{+}$ $(1^{+})$ $c\bar{s}$ state to the $DK$ 
$(D^{\ast}K)$ threshold plays an important dynamical role in lowering the bare 
mass to the observed value.

In a recent lattice study of the $D_{s0}^{\ast}(2317)$ 
meson~\cite{Mohler:2013rwa}, good agreement with the experimental mass is found 
when operators for $DK$ scattering states are included. An extended version of 
the work performed in~\cite{Mohler:2013rwa} was presented in 
Ref.~\cite{Lang:2014yfa}. They study the $J^{P}=0^{+}$, $1^{+}$ and $2^{+}$ 
charmed-strange mesons incorporating the effect of nearby $DK$ and $D^{\ast}K$ 
thresholds. The $D^{\ast}K$ threshold is incorporated only as an $S$-wave 
channel in the lattice QCD computations. However, the $D$-wave $D^{\ast}K$ 
channel could play an important role in the $1^{+}$ $c\bar{s}$ sector, in 
particular for the $j_{q}^{P}=3/2^{+}$ $D_{s1}$ meson. Moreover, despite the 
significant progress made by lattice calculations incorporating $DK$ and 
$D^{\ast}K$ thresholds, no statement can be made about the probabilities of the 
different Fock components in the physical state.

The authors of Ref.~\cite{Torres:2014vna} re-analyzed the lattice spectrum 
obtained in Refs.~\cite{Mohler:2013rwa, Lang:2014yfa} using the auxiliary 
potential method and a reformulation (valid only for $S$-waves scattering 
amplitudes) of the Weinberg compositeness condition~\cite{Weinberg:1965zz, 
Baru:2003qq} to determine the amount of $DK$ and $D^{\ast}K$ components in the 
respective wave functions of $D_{s0}^{\ast}(2317)$ and $D_{s1}(2460)$ mesons. 
They found that the $D_{s0}^{\ast}(2317)$ meson is made by $(72\pm13\pm5)\%$ of 
$DK$ component whereas the $D_{s1}(2460)$ contains a $(57\pm21\pm6)\%$ of 
$D^{\ast}K$.

In this paper, we study the low-lying $P$-wave charmed-strange mesons using a 
nonrelativistic constituent quark model in which quark-antiquark and 
meson-meson degrees of freedom are incorporated. The constituent quark model 
(CQM) was proposed in Ref.~\cite{Vijande:2004he} (see 
references~\cite{Valcarce:2005em} and~\cite{Segovia:2013wma} for reviews). 
This model successfully describes hadron phenomenology and hadronic 
reactions~\cite{Fernandez:1992xs, Garcilazo:2001md, Vijande:2004at} and has 
recently been applied to (non)conventional hadrons containing heavy quarks 
(see, for instance, Refs.~\cite{Segovia:2011zza, Ortega:2012cx, Segovia:2013kg, 
Ortega:2014eoa, Ortega:2014fha, Segovia:2014mca, Segovia:2016xqb}).

Within our approach, the coupling between the quark-antiquark and meson-meson 
sectors requires the creation of a light quark-antiquark pair. The associated 
operator should be similar to the one which describes the open-flavour meson 
strong decays, namely the $^{3}P_{0}$ transition 
operator~\cite{LeYaouanc:1972ae}. 

Our calculation allows to introduce the coupling with the $D$-wave $D^{\ast}K$ 
channel in the $1^{+}$ $c\bar{s}$ sector and the computation of the 
probabilities associated with the different Fock components of the physical 
state, features which cannot be addressed nowadays by any other theoretical 
approach by itself.

This manuscript is arranged as follows. In Sec.~\ref{sec:theory} we 
describe the main properties of our theoretical formalism giving details 
about the approaches used to describe the quark-antiquark sector, the 
meson-meson sector and the coupling between them. Section~\ref{sec:results} is 
devoted to present our results for the $D_{s0}^{\ast}(2317)$, $D_{s1}(2460)$, 
$D_{s1}(2536)$ and $D_{s2}^{\ast}(2573)$ mesons. We finish summarizing and 
giving some conclusions in Sec.~\ref{sec:epilogue}.

%%%%%%%%%%%%%%%%%%%%%%%%%%%%%%%%%%%%%%%%%%%%%%%%%%%%%%%%%%%%%%%%%%%%%%%%%%%%%%%
%%%%%%%%%%%%%%%%%%%%%%%%%%%%%%%%%%%%%%%%%%%%%%%%%%%%%%%%%%%%%%%%%%%%%%%%%%%%%%%

\section{THEORETICAL FORMALISM}
\label{sec:theory} 

\subsection{Naive quark model}
\label{subsec:quarkmodel}

Constituent light quark masses and Goldstone-boson exchanges, which are 
consequences of dynamical chiral symmetry breaking in Quantum Chromodynamics 
(QCD), together with the perturbative one-gluon exchange (OGE) and a 
nonperturbative confining interactions are the main pieces of our constituent 
quark model~\cite{Vijande:2004he, Segovia:2013wma}.

A simple Lagrangian invariant under chiral transformations can be written in 
the following form~\cite{Diakonov:2002fq}
\begin{equation}
{\mathcal L} = \bar{\psi}(i\, {\slash\!\!\! \partial} 
-M(q^{2})U^{\gamma_{5}})\,\psi  \,,
\end{equation}
where $M(q^2)$ is the dynamical (constituent) quark mass and $U^{\gamma_5} = 
e^{i\lambda _{a}\phi ^{a}\gamma _{5}/f_{\pi}}$ is the matrix of Goldstone-boson 
fields that can be expanded as
\begin{equation}
U^{\gamma _{5}} = 1 + \frac{i}{f_{\pi}} \gamma^{5} \lambda^{a} \pi^{a} - 
\frac{1}{2f_{\pi}^{2}} \pi^{a} \pi^{a} + \ldots
\end{equation}
The first term of the expansion generates the constituent quark mass while the
second gives rise to a one-boson exchange interaction between quarks. The
main contribution of the third term comes from the two-pion exchange which
has been simulated by means of a scalar-meson exchange potential.

In the heavy quark sector chiral symmetry is explicitly broken and 
Goldstone-boson exchanges do not appear. However, it constrains the model 
parameters through the light-meson phenomenology~\cite{Segovia:2008zza} and 
provides a natural way to incorporate the pion exchange interaction in the 
molecular dynamics.

It is well known that multi-gluon exchanges produce an attractive linearly 
rising potential proportional to the distance between infinite-heavy quarks. 
However, sea quarks are also important ingredients of the strong interaction 
dynamics that contribute to the screening of the rising potential at low 
momenta and eventually to the breaking of the quark-antiquark binding 
string~\cite{Bali:2005fu}. Our model tries to mimic this behaviour using the 
following expression:
\begin{equation}
V_{\rm CON}(\vec{r}\,)=\left[-a_{c}(1-e^{-\mu_{c}r})+\Delta \right] 
(\vec{\lambda}_{q}^{c}\cdot\vec{\lambda}_{\bar{q}}^{c}) \,,
\label{eq:conf}
\end{equation}
where $a_{c}$ and $\mu_{c}$ are model parameters. At short distances this 
potential presents a linear behaviour with an effective confinement strength,
$\sigma=-a_{c}\,\mu_{c}\,(\vec{\lambda}^{c}_{i}\cdot \vec{\lambda}^{c}_{j})$,
while it becomes constant at large distances. This type of potential shows a
threshold defined by
\begin{equation}
V_{\rm thr}=\{-a_{c}+ \Delta\}(\vec{\lambda}^{c}_{i}\cdot
\vec{\lambda}^{c}_{j}).
\end{equation}
No quark-antiquark bound states can be found for energies higher than this
threshold. The system suffers a transition from a colour string configuration
between two static colour sources into a pair of static mesons due to the
breaking of the colour flux-tube and the most favoured subsequent decay into
hadrons.

The OGE potential is generated from the vertex Lagrangian
\begin{equation}
{\mathcal L}_{qqg} = i\sqrt{4\pi\alpha_{s}} \, \bar{\psi} \gamma_{\mu} 
G^{\mu}_{c} \lambda^{c} \psi,
\label{Lqqg}
\end{equation}
where $\lambda^{c}$ are the $SU(3)$ colour matrices, $G^{\mu}_{c}$ is the
gluon field and $\alpha_{s}$ is the strong coupling constant. The scale 
dependence of $\alpha_{s}$ in order to get in our approach a consistent 
description of light, strange and heavy mesons can be found in 
Ref.~\cite{Vijande:2004he}. 

To improve the description of the open-flavour mesons, we follow the proposal 
of Ref.~\cite{Lakhina:2006fy} and include one-loop corrections to the OGE 
potential as derived by Gupta {\it et al.}~\cite{Gupta:1981pd}. These 
corrections show a spin-dependent term which affects only mesons with different 
flavour quarks. The net result is a quark-antiquark interaction that can be 
written as:
\begin{equation}
V(\vec{r}_{ij})=V_{\rm OGE}(\vec{r}_{ij})+V_{\rm CON}(\vec{r}_{ij})+V_{\rm 
OGE}^{\rm 1-loop}(\vec{r}_{ij}),
\end{equation}
where $V_{\rm OGE}$ and $V_{\rm CON}$ have been introduced above and the 
$V_{\rm OGE}^{\rm 1-loop}$ term is the one-loop correction to the OGE potential 
that  contains central, tensor and spin-orbit contributions whose particular 
expressions implemented in our quark model can be found in 
Ref.~\cite{Segovia:2012yh}.

Explicit expressions for all the potentials and the value of the model 
parameters can be found in Ref.~\cite{Vijande:2004he}, updated in 
Refs.~\cite{Segovia:2008zz}. Meson eigenenergies and eigenstates are obtained 
by solving the Schr\"odinger equation using the Gaussian Expansion 
Method~\cite{Hiyama:2003cu} which provides enough accuracy and it simplifies 
the subsequent evaluation of the needed matrix elements.

Following Ref.~\cite{Hiyama:2003cu}, we employ Gaussian trial functions with
ranges in geometric progression. This enables the optimization of ranges
employing a small number of free parameters. Moreover, the geometric
progression is dense at short distances, so that it enables the description of
the dynamics mediated by short range potentials. The fast damping of the
Gaussian tail does not represent an issue, since we can choose the maximal
range much larger than the hadronic size.

%%%%%%%%%%%%%%%%%%%%%%%%%%%%%%%%%%%%%%%%%%%%%%%%%%%%%%%%%%%%%%%%%%%%%%%%%%%%%%%

\subsection{Coupled-channel quark model}
\label{subsec:coupledchannel} 

The quark-antiquark bound state can be strongly influenced by nearby 
multiquark channels. In this work, we follow Ref.~\cite{Ortega:2010qq} to study 
this effect in the spectrum of the charmed-strange mesons and thus we need to 
assume that the hadronic state is given by
\begin{equation} 
| \Psi \rangle = \sum_\alpha c_\alpha | \psi_\alpha \rangle
+ \sum_\beta \chi_\beta(P) |\phi_A \phi_B \beta \rangle,
\label{ec:funonda}
\end{equation}
where $|\psi_\alpha\rangle$ are $c\bar{s}$ eigenstates of the two-body 
Hamiltonian, $\phi_{M}$ are wave functions associated with the $A$ and $B$ 
mesons, $|\phi_A \phi_B \beta \rangle$ is the two meson state with $\beta$ 
quantum numbers coupled to total $J^{PC}$ quantum numbers and $\chi_\beta(P)$ 
is the relative wave function between the two mesons in the molecule. When we 
solve the four-body problem we also use the Gaussian Expansion Method (GEM) of 
the $q\bar{q}$ wave functions obtained from the solution of the two-body 
problem. This procedure allows us to introduce in a variational way possible 
distortions of the two-body wave function within the molecule. To derive the 
meson-meson interaction from the $q\bar{q}$ interaction we use the Resonating 
Group Method (RGM)~\cite{Tang:1978zz}.

The coupling between the quark-antiquark and meson-meson sectors requires the
creation of a light quark pair. The operator associated with this process 
should describe also the open-flavour meson strong decays and is given 
by~\cite{Segovia:2012cd}
\begin{equation}
\begin{split}
T =& -\sqrt{3} \, \sum_{\mu,\nu}\int d^{3}\!p_{\mu}d^{3}\!p_{\nu}
\delta^{(3)}(\vec{p}_{\mu}+\vec{p}_{\nu})\frac{g_{s}}{2m_{\mu}}\sqrt{2^{5}\pi}
\,\times \\
&
\times \left[\mathcal{Y}_{1}\left(\frac{\vec{p}_{\mu}-\vec{p}_{\nu}}{2}
\right)\otimes\left(\frac{1}{2}\frac{1}{2}\right)1\right]_{0}a^{\dagger}_{\mu}
(\vec{p}_{\mu})b^{\dagger}_{\nu}(\vec{p}_{\nu}) \,.
\label{eq:Otransition2}
\end{split}
\end{equation}
where $\mu$ $(\nu)$ are the spin, flavour and colour quantum numbers of the
created quark (antiquark). The spin of the quark and antiquark is coupled to
one. The ${\cal Y}_{lm}(\vec{p}\,)=p^{l}Y_{lm}(\hat{p})$ is the solid harmonic
defined in function of the spherical harmonic. We fix the relation of $g_{s}$ 
with the dimensionless constant giving the strength of the quark-antiquark pair 
creation from the vacuum as $\gamma=g_{s}/2m$, being $m$ the mass of the 
created quark (antiquark).

It is important to emphasize here that the $^{3}P_{0}$ model depends only on 
one parameter, the strength $\gamma$ of the decay interaction. Some attempts 
have been done to find possible dependences of the vertex parameter $\gamma$, 
see~\cite{Ferretti:2013vua} and references therein. In 
Ref.~\cite{Segovia:2012cd} we performed a successful fit to the decay widths of 
the mesons which belong to charmed, charmed-strange, hidden charm and hidden 
bottom sectors and elucidated the dependence on the mass scale of the 
$^{3}P_{0}$ free parameter $\gamma$. Further details about the global fit can 
be found in Ref.~\cite{Segovia:2012cd}. The running of the strength $\gamma$ of 
the $^{3}P_{0}$ decay model is given by
\begin{equation}
\gamma(\mu) = \frac{\gamma_{0}}{\log\left(\frac{\mu}{\mu_{0}}\right)},
\label{eq:fitgamma}
\end{equation}
where $\gamma_{0}$ and $\mu_{0}$ are parameters, whereas $\mu$ is the reduced 
mass of the quark-antiquark in the decaying meson. The value of $\gamma$ that 
we are using in this work is the one corresponding to the charmed-strange 
sector: $\gamma=0.38$.

From the operator in Eq.~(\ref{eq:Otransition2}), we define the transition
potential $h_{\beta \alpha}(P)$ within the $^{3}P_{0}$ model
as~\cite{Kalashnikova:2005ui} 
\begin{equation}
\langle \phi_{M_1} \phi_{M_2} \beta | T | \psi_\alpha \rangle =
P \, h_{\beta \alpha}(P) \,\delta^{(3)}(\vec P_{\rm cm})\,,
\label{Vab}
\end{equation}
where $P$ is the relative momentum of the two-meson state.

The usual version of the $^{3}P_{0}$ model gives vertices that are too hard 
specially when we work at high momenta. Following the suggestion of 
Ref.~\cite{Morel:2002vk}, we use a momentum dependent form factor to truncate 
the vertex as
\begin{equation}
\label{Vab mod}
h_{\beta \alpha}(P)\to h_{\beta \alpha}(P)\times 
e^{-\frac{P^2}{2\Lambda^2}} \,,
\end{equation}
where $\Lambda=0.84\,{\rm GeV}$ is the value used herein.

Adding the coupling with charmed-strange states we end-up with the 
coupled-channels equations
\begin{equation}
\begin{split}
&
c_\alpha M_\alpha +  \sum_\beta \int h_{\alpha\beta}(P) \chi_\beta(P)P^2 dP = E
c_\alpha\,, \\
&
\sum_{\beta}\int H_{\beta'\beta}(P',P)\chi_{\beta}(P) P^2 dP + \\
&
\hspace{2.50cm} + \sum_\alpha h_{\beta'\alpha}(P') c_\alpha = E
\chi_{\beta'}(P')\,,
\label{ec:Ec-Res}
\end{split}
\end{equation}
where $M_\alpha$ are the masses of the bare $c\bar{s}$ mesons and 
$H_{\beta'\beta}$ is the RGM Hamiltonian for the two-meson states obtained from
the $q\bar{q}$ interaction. Solving the coupling with the $c\bar{s}$ states, we
arrive to a Schr\"odinger-type equation
\begin{equation}
\begin{split}
\sum_{\beta} \int \big( H_{\beta'\beta}(P',P) + &
V^{\rm eff}_{\beta'\beta}(P',P) \big) \times \\
&
\times \chi_{\beta}(P) {P}^2 dP = E \chi_{\beta'}(P'),
\label{ec:Ec1}
\end{split}
\end{equation}
where
\begin{equation}
V^{\rm eff}_{\beta'\beta}(P',P;E)=\sum_{\alpha}\frac{h_{\beta'\alpha}(P')
h_{\alpha\beta}(P)}{E-M_{\alpha}}.
\end {equation}

Finally, let us mention that this version of the coupled-channel quark model 
has been applied extensively to the study of XYZ states (see, for instance, 
Ref.~\cite{Ortega:2012rs}).

%%%%%%%%%%%%%%%%%%%%%%%%%%%%%%%%%%%%%%%%%%%%%%%%%%%%%%%%%%%%%%%%%%%%%%%%%%%%%%%
%%%%%%%%%%%%%%%%%%%%%%%%%%%%%%%%%%%%%%%%%%%%%%%%%%%%%%%%%%%%%%%%%%%%%%%%%%%%%%%

\section{RESULTS}
\label{sec:results}

Table~\ref{tab:1loopDmesons} shows the masses of the low-lying $P$-wave 
charmed-strange mesons predicted by the naive quark model. One can see our 
results taking into account the one-gluon exchange potential $(\alpha_{s})$ and 
including its one-loop corrections $(\alpha_{s}^{2})$. The experimental data 
are taken from the Review of Particle Physics (RPP)~\cite{Agashe:2014kda}.

\begin{table}[!t]
\caption{\label{tab:1loopDmesons} Masses, in MeV, of the low-lying $P$-wave 
charmed-strange mesons predicted by the constituent quark model $(\alpha_{s})$
and those including one-loop corrections to the one-gluon exchange potential
$(\alpha_{s}^{2})$. Experimental data are taken from 
Ref.~\cite{Agashe:2014kda}.}
\begin{ruledtabular}
\begin{tabular}{llccc}
State & $J^{P}$ & The. $(\alpha_{s})$ & The. $(\alpha_{s}^{2})$ & Exp. \\
\hline
$D_{s0}^{\ast}(2317)$ & $0^{+}$ & $2511$ & $2383$ & $2318.0\pm1.0$ \\
$D_{s1}(2460)$        & $1^{+}$ & $2593$ & $2570$ & $2459.6\pm0.9$ \\
$D_{s1}(2536)$        & $1^{+}$ & $2554$ & $2560$ & $2535.18\pm0.24$ \\
$D_{s2}^{\ast}(2573)$ & $2^{+}$ & $2592$ & $2609$ & $2571.9\pm0.8$ \\
\end{tabular}
\end{ruledtabular}
\end{table}

The naive quark model predicts masses for the $D_{s0}^{\ast}(2317)$ and 
$D_{s1}(2460)$ mesons much higher than the experimental values. In fact, one 
can conclude from Table~\ref{tab:1loopDmesons} that the $j_q^P=1/2^{+}$ 
and $3/2^{+}$ doublets are predicted to be almost degenerated within the naive 
quark model. The state assigned to the $D_{s0}^{\ast}(2317)$ is very sensitive 
to the $1$-loop corrections of the OGE potential which bring its mass closer to 
the experimental one. This effect could explain part of its lower mass but it 
is not enough because our theoretical state is still above the $DK$ threshold. 
The mass associated with the $D_{s1}(2460)$ meson is roughly insensitive to the 
spin-dependent 1-loop corrections of the OGE potential.

One can conclude from above that possible canonical $c\bar{s}$ descriptions of 
the $D_{s0}^{\ast}(2317)$ and $D_{s1}(2460)$ mesons seem to fail when model 
parameters are kept to describe other quark sectors. From the conclusions of 
recent lattice-regularised QCD computations~\cite{Mohler:2013rwa, 
Lang:2014yfa}, the coupling of the $J^{P}=0^{+}$ $(1^{+})$ $c\bar{s}$ state to 
the $DK$ $(D^{\ast}K)$ threshold appears as a possible mechanism to bring our 
theoretical masses to the experimental values. 

HQS predicts that the members of the $j_q^P=1/2^{+}$ doublet 
($D_{s0}^{\ast}(2317)$ and $D_{s1}(2460)$) couple equally to their respective 
$DK$ and $D^{\ast}K$ thresholds~\cite{JuanNieves}. Moreover, because these 
states have the same mass in the limit $m_{Q}\to \infty$, the potentially 
generated mass shifts depend only on the energy difference between the bare 
$c\bar{s}$ state and the open-flavoured threshold. This would give a mass shift 
larger for the $1^{+}$ $c\bar{s}$ state than for the $0^{+}$ one, which is 
contrary to the experimental situation. The $1$-loop corrections of the OGE 
potential solve this issue and provide appropriate bare states whose mass 
shifts due to the continuum go in accordance with experiment.

%%%%%%%%%%%%%%%%%%%%%%%%%%%%%%%%%%%%%%%%%%%%%%%%%%%%%%%%%%%%%%%%%%%%%%%%%%%%%%%

\begin{table}[!t]
\caption{\label{tab:Ds2317} Values of $m_{D_{s0}^{\ast}(2317)} - 
m_{\overline{1S}}$, in MeV, predicted by our quark model and lattice 
QCD~\cite{Mohler:2013rwa} taking into account only quark-antiquark degrees of 
freedom and also coupling with the $DK$ threshold. The 
$m_{D_{s0}^{\ast}(2317)}$ is the mass of the $D_{s0}^{\ast}(2317)$ state and 
$m_{\overline{1S}} = 1/4 (m_{D_{s}} + 3m_{D_{s}^{\ast}})$ is the spin-averaged 
ground state mass. We compare with the experimental data taken form 
Ref.~\cite{Agashe:2014kda}.}
\begin{ruledtabular}
\begin{tabular}{lcccc}
$D_{s0}^{\ast}(2317)$ & CQM & LQCD (1) & LQCD (2) & Exp. \\
\hline
$q\bar{q}$    & $309.0$ & $274.7\pm15.8$ & $320.4\pm21.3$ & $241.7\pm1.1$ \\
$q\bar{q}+DK$ & $249.6$ & $254.4\pm4.9$  & $245.0\pm15.5$ & $241.7\pm1.1$ \\
$V\to\infty$  & -       & $287.2\pm5.8$  & $266.0\pm16.5$ & $241.7\pm1.1$ \\
\end{tabular}
\end{ruledtabular}
\end{table}

\begin{figure*}[!t]
\begin{center}
\epsfig{figure=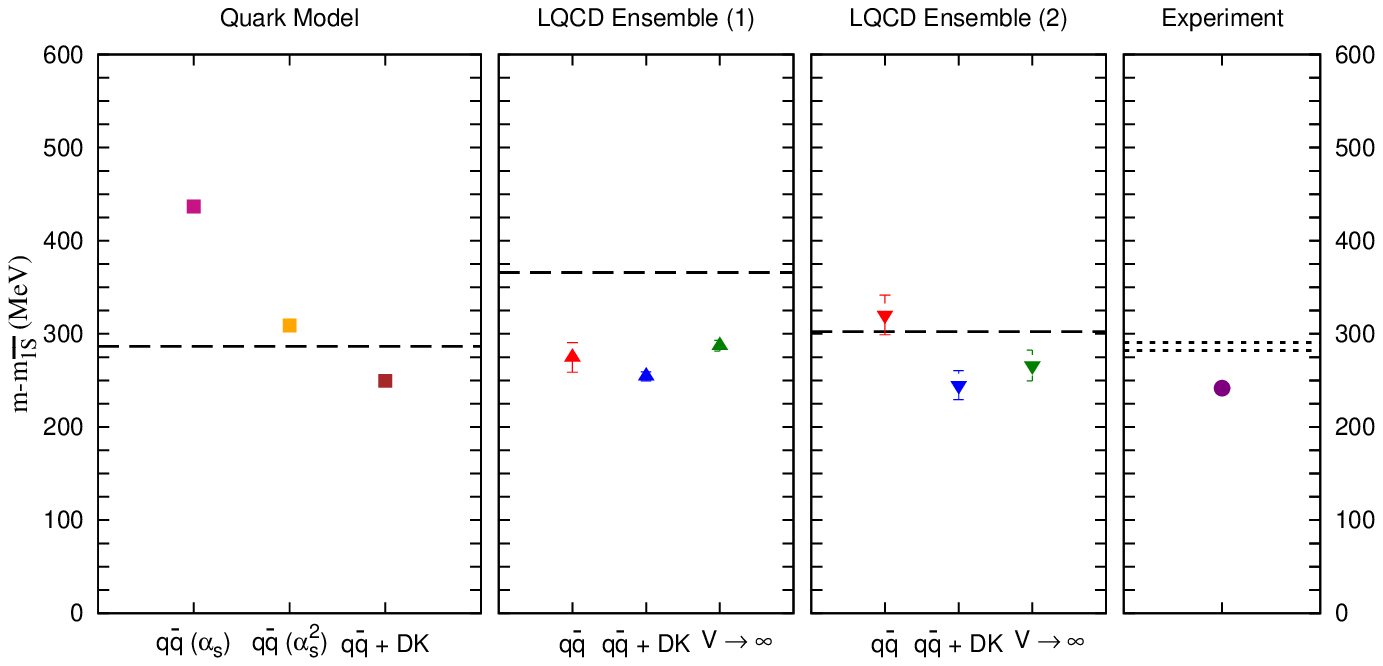,height=0.25\textheight,width=0.80\textwidth}
\caption{\label{fig:Ds2317} Energy levels from constituent quark model (CQM), 
from Lattice QCD~\cite{Mohler:2013rwa} using Ensemble~(1) and Ensemble~(2), and 
from experiment~\cite{Agashe:2014kda}. We show, for CQM results, the 
quark-antiquark value taking into account the one-gluon exchange potential 
$(\alpha_{s})$, including its one-loop corrections $(\alpha_{s}^{2})$ and 
coupling with the $DK$ threshold. For the lattice QCD results, in each 
ensemble, we show values with just a $q\bar{q}$ interpolator basis and with a 
combined basis of $q\bar{q}$ and $DK$ interpolating fields. The value of the 
bound $D_{s0}^{\ast}(2317)$ state position in the infinite volume limit $V\to 
\infty$ is obtained by an analytical continuation of the scattering amplitude 
combined with L\"uscher's finite volume method. The dashed lines represent the 
threshold for $DK$ in each approach and the dotted lines are the thresholds for 
$D^{0}K^{+}$ and $D^{+}K^{0}$ in experiment.}
\end{center}
\end{figure*}

\begin{table}[!t]
\caption{\label{tab:Ds2317P} Mass, in MeV, and probabilities of the different 
Fock components, in \%, of the ${D_{s0}^{\ast}(2317)}$ state.}
\begin{ruledtabular}
\begin{tabular}{ccccc}
State & Mass & ${\cal P}[q\bar{q}\,(^{3}P_{0})]$ & ${\cal P}[DK(S-wave)]$ \\
\hline
$D_{s0}^{\ast}(2317)$ & $2323.7$ & $66.3\%$ &  $33.7\%$ \\
\end{tabular}
\end{ruledtabular}
\end{table}

\subsection{The dressed $D_{s0}^{\ast}(2317)$ meson}
\label{subsec:0p}

Table~\ref{tab:Ds2317} and Fig.~\ref{fig:Ds2317} compare our results for the 
$D_{s0}^{\ast}(2317)$ mass with the lattice QCD study of 
Ref.~\cite{Mohler:2013rwa} and with experiment~\cite{Agashe:2014kda}. Instead 
of the $D_{s0}^{\ast}(2317)$ itself, following the lattice study, we compare 
the values of $m_{D_{s0}^{\ast}(2317)} - m_{\overline{1S}}$, where 
$m_{\overline{1S}} = 1/4 (m_{D_{s}}+3m_{D_{s}^{\ast}})$ is the spin-averaged 
ground state mass. Note that the lattice value of the $D_{s0}^{\ast}(2317)$ 
bound state position in the infinite volume limit ($V\to\infty$) is obtained by 
an analytical continuation of the scattering amplitude combined with 
L\"uscher's finite volume method~\cite{Mohler:2013rwa, Lang:2014yfa}. In 
Fig.~\ref{fig:Ds2317}, the dashed lines represent the threshold for $DK$ in the 
different approaches and the dotted lines are the thresholds for $D^{0}K^{+}$ 
and $D^{+}K^{0}$ in experiment.

The mass of the $D_{s0}^{\ast}(2317)$ state obtained using the naive quark 
model and without the $1$-loop spin corrections to the OGE potential is much 
higher than the experimental value. In this case, the $m_{D_{s0}^{\ast}(2317)} 
- m_{\overline{1S}}=437\,{\rm MeV}$ is almost twice the experimental value. As 
we have discussed previously, the mass associated to the $D_{s0}^{\ast}(2317)$ 
state is very sensitive to the $\alpha_{s}^{2}$-corrections of the OGE 
potential. This effect brings down the $m_{D_{s0}^{\ast}(2317)} - 
m_{\overline{1S}}$ splitting to $309\,{\rm MeV}$, which is now only $30\%$ 
higher than the experimental figure. However, as one can see in 
Fig.~\ref{fig:Ds2317}, the hypothetical $D_{s0}^{\ast}(2317)$ would be above 
the $DK$ threshold and thus would decay into this final channel in an $S$-wave 
making the state wider than the observed one. The mass-shift due to the 
$\alpha_{s}^{2}$-corrections allows that the $0^{+}$ state be close to the $DK$ 
threshold. This makes the $DK$ coupling a relevant dynamical mechanism in the 
formation of the $D_{s0}^{\ast}(2317)$ bound state. When we couple the $0^{+}$ 
$c\bar{s}$ ground state with the $DK$ threshold, the splitting 
$m_{D_{s0}^{\ast}(2317)} - m_{\overline{1S}}=249.6\,{\rm MeV}$ is in good 
agreement with experiment.

The lattice QCD simulation is done on two very different ensembles of gauge 
configurations: Ensemble~(1) with $2$ dynamical quarks, a pion mass of 
$266\,{\rm MeV}$ and a lattice size of $16^3\times 32$; and Ensemble~(2) with 
$2+1$ dynamical quarks, a pion mass of $156\,{\rm MeV}$ and a lattice size of 
$32^3\times 64$. One can see in Fig.~\ref{fig:Ds2317} that the outcome from 
lattice simulations depends somewhat delicately on the pion mass even for very 
low masses. For the case of largest pion mass and only quark-antiquark 
interpolators, unlike previous lattice simulations, the $D_{s0}^{\ast}$ appears 
below the $DK$ threshold with a $m_{D_{s0}^{\ast}(2317)}-m_{\overline{1S}}$ in 
reasonable agreement with the experimental value; the inclusion of $DK$ 
interpolators produce a little effect in this case (see LQCD (1) results in 
Table~\ref{tab:Ds2317} and Fig.~\ref{fig:Ds2317}). In a near to physical light 
quark mass simulation (LQCD (2)), the $D_{s0}^{\ast}(2317)$ is above $DK$ 
threshold when only quark-antiquark interpolators are included. In this case, 
the combination of $q\bar{q}$ and $DK$ lattice interpolating fields is crucial 
in order to get agreement with experiment. Finally, Table~\ref{tab:Ds2317} and 
Fig.~\ref{fig:Ds2317} show also the physical extrapolation of the 
$m_{D_{s0}^{\ast}(2317)} - m_{\overline{1S}}$ splitting in both ensembles. This 
value agrees with experiment and with our result when we incorporate the 
coupling of the $DK$ threshold to the $0^{+}$ $c\bar{s}$ state.

We turn now to discuss the probabilities of the different Fock components in 
the physical state. Lattice QCD studies~\cite{Mohler:2013rwa, Lang:2014yfa} are 
only able to remark that both quark-antiquark and meson-meson lattice 
interpolating fields have non-vanishing overlaps with the physical state. Our 
wave function probabilities are given in Table~\ref{tab:Ds2317P} which reflects 
that the $D_{s0}^{\ast}(2317)$ meson is mostly of quark-antiquark nature. This 
is in agreement with the fact that lattice-regularised QCD computations observe 
this state even with only $q\bar{q}$ interpolators (see Fig.~\ref{fig:Ds2317}). 
However is markedly in contrast with the $70\%$ of $DK$ obtained by 
Ref.~\cite{Torres:2014vna} in the analysis of the lattice data of 
Refs.~\cite{Mohler:2013rwa, Lang:2014yfa}.

In our model the probability of the $DK$ state depends basically on three 
quantities: the bare meson mass, the $^3P_0$ coupling constant and the residual 
$DK$ interaction. Obviously, as neither of the three are observables, they can 
take different values depending on the dynamics, making the results, and hence 
the $DK$ probability, model dependent.

In this paper we have constrained the mentioned parameters by reproducing other 
observable quantities like strong decays~\cite{Segovia:2012cd} (the $^3P_0$ 
coupling constant), charmonium spectrum~\cite{Segovia:2008zz} (the bare mass) 
and $NN$ and $p\bar p$ interactions~\cite{Entem:2000mq, Entem:2006dt} (the 
$DK$ residual interaction).

To check the uncertainties of the model, we have varied the value of the bare 
$c\bar s$ mass and the $^3P_0$ coupling $\gamma$ keeping the mass of the 
physical state to the experimental value. As expected in a model where the 
$DK$ interaction is smaller than the effective interaction due to the coupling 
with the $c\bar s$ state, we obtain that the probability of the $c\bar s$ 
component grows as the bare mass approaches the physical mass. To reproduce the 
scenario presented in Ref.~\cite{Torres:2014vna} we would need a stronger 
residual $DK$ interaction incompatible with the limits of the model. However, 
other dynamics are possible in quark models~\cite{Dai:2003dz}; its analysis 
would be interesting in order to explore the possible convergence to the result 
of Ref.~\cite{Torres:2014vna} but this task goes beyond the scope of the 
present work.

%%%%%%%%%%%%%%%%%%%%%%%%%%%%%%%%%%%%%%%%%%%%%%%%%%%%%%%%%%%%%%%%%%%%%%%%%%%%%%%

\begin{figure*}[!t]
\begin{center}
\epsfig{figure=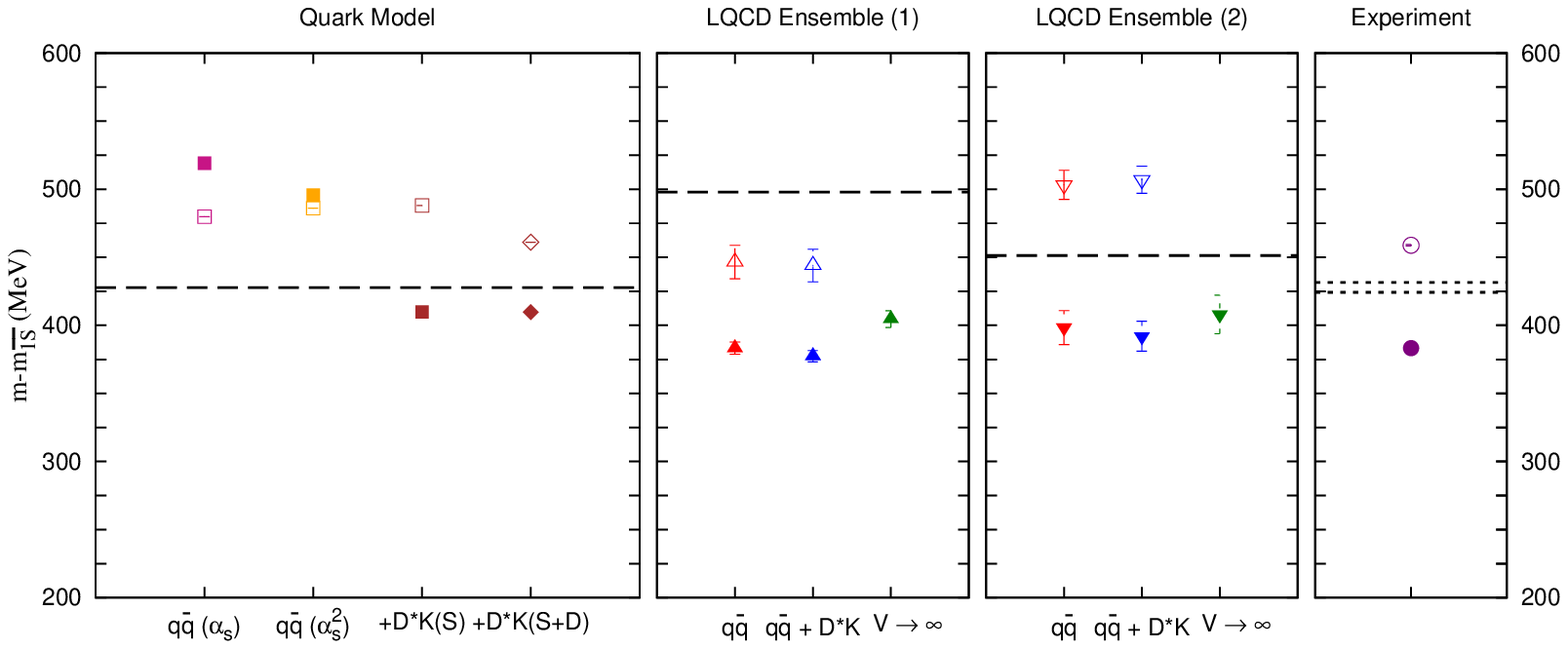,height=0.25\textheight,width=0.80\textwidth}
\caption{\label{fig:Ds24602536} Energy levels from constituent quark model 
(CQM), from Lattice QCD~\cite{Lang:2014yfa} using Ensemble~(1) and 
Ensemble~(2), and from experiment~\cite{Agashe:2014kda}. We show, for CQM 
results, the quark-antiquark value taking into account the one-gluon exchange 
potential $(\alpha_{s})$, including its one-loop corrections $(\alpha_{s}^{2})$ 
and coupling with the $D^{\ast}K$ threshold in $S$- and $D$-wave. For the 
lattice QCD results, in each case, we show values with just a $q\bar{q}$ 
interpolator basis and with a combined basis of $q\bar{q}$ and $D^{\ast}K$ 
interpolating fields. Remember that in the lattice QCD computations the 
$D^{\ast}K$ threshold is coupled only in an $S$-wave. The value of the bound 
$D_{s1}(2460)$ state position in the infinite volume limit $V\to \infty$ is 
obtained by an analytical continuation of the scattering amplitude combined 
with L\"uscher's finite volume method. This method has not been used for the 
$D_{s1}(2536)$ meson. The dashed lines represent the threshold for $D^{\ast}K$ 
in each approach and the dotted lines are the thresholds for $D^{\ast0}K^{+}$ 
and $D^{\ast+}K^{0}$ in experiment.}
\end{center}
\end{figure*}

\begin{table}[!t]
\caption{\label{tab:Ds24602536} Values of $m_{D_{s1}(2460)} - 
m_{\overline{1S}}$ and $m_{D_{s1}(2536)} - m_{\overline{1S}}$, in MeV, 
predicted by our quark model and lattice QCD~\cite{Lang:2014yfa} taking into 
account only quark-antiquark degrees of freedom and also coupling with the 
$D^{\ast}K$ threshold. The $m_{D_{s1}(2460)}$ and $m_{D_{s1}(2536)}$ are the 
masses of the $D_{s1}(2460)$ and $D_{s1}(2536)$ states and $m_{\overline{1S}} = 
1/4 (m_{D_{s}} + 3m_{D_{s}^{\ast}})$ is the spin-averaged ground state mass. We 
compare with the experimental data taken form Ref.~\cite{Agashe:2014kda}.}
%\begin{ruledtabular}
\scalebox{0.95}{\begin{tabular}{lcccc}
\hline\hline
$D_{s1}(2460)$ & CQM & LQCD (1) & LQCD (2) & Exp. \\
\hline
$q\bar{q}$                & $495.6$ & $383.3\pm4.5$ & $398.4\pm12.5$ & 
$383.3\pm1.0$ \\
$q\bar{q}+D^{\ast}K(S)$   & $409.9$ & $377.4\pm4.2$ & $392.0\pm11.0$ & 
$383.3\pm1.0$ \\
$V\to\infty$              & -       & $404.6\pm6.2$ & $408.0\pm14.2$ & 
$383.3\pm1.0$ \\
$q\bar{q}+D^{\ast}K(S+D)$ & $409.8$ & -             & -              & 
$383.3\pm1.0$ \\
\hline
\hline
$D_{s1}(2536)$ & CQM & LQCD (1) & LQCD (2) & Exp. \\
\hline
$q\bar{q}$                & $486.0$ & $446.5\pm12.3$ & $503.2\pm10.7$ & 
$458.9\pm0.5$ \\
$q\bar{q}+D^{\ast}K(S)$   & $488.0$ & $444.0\pm12.0$ & $507.0\pm10.0$ & 
$458.9\pm0.5$ \\
$V\to\infty$              & -       & -       & -       & $458.9\pm0.5$ \\
$q\bar{q}+D^{\ast}K(S+D)$ & $461.1$ & -       & -       & $458.9\pm0.5$ \\
\hline\hline
\end{tabular}}
%\end{ruledtabular}
\end{table}

\begin{table*}[!t]
\caption{\label{tab:Ds2460P} Mass and decay width, in MeV; and probabilities of 
the different Fock components, in \%, of the ${D_{s1}(2460)}$ and 
$D_{s1}(2536)$ states. Results with and without coupling of the $D$-wave 
$D^{\ast}K$ channel are listed.}
\begin{ruledtabular}
\begin{tabular}{ccccccc}
State & Mass & Width & ${\cal P}[q\bar{q}\,(^{1}P_{1})]$ & ${\cal 
P}[q\bar{q}\,(^{3}P_{1})]$ & ${\cal P}[D^{\ast}K(S-wave)]$ & ${\cal 
P}[D^{\ast}K(D-wave)]$\\
\hline
$D_{s1}(2460)$ & $2484.0$ & $0.00$ & $12.9\%$ & $32.8\%$ & $54.3\%$ & - \\
$D_{s1}(2536)$ & $2562.1$ & $0.22$ & $34.4\%$ & $15.8\%$ & $49.8\%$ & - \\
\hline
$D_{s1}(2460)$ & $2484.0$ & $0.00$ & $12.1\%$ & $33.6\%$ & $54.1\%$ & $0.2\%$ \\
$D_{s1}(2536)$ & $2535.2$ & $0.56$ & $31.9\%$ & $14.5\%$ & $16.8\%$ & $36.8\%$ 
\\
\end{tabular}
\end{ruledtabular}
\end{table*}

\subsection{The dressed $D_{s1}(2460)$ and $D_{s1}(2536)$ mesons}
\label{subsec:1p}

Table~\ref{tab:Ds24602536} and Fig.~\ref{fig:Ds24602536} compare our results 
for the mass of the first two $J^{P}=1^{+}$ charmed-strange states with the 
lattice QCD study of Ref.~\cite{Lang:2014yfa} and with 
experiment~\cite{Agashe:2014kda}. Instead of the masses themselves, following 
the lattice study, we compare their difference with respect the spin-averaged 
ground state mass, $m_{\overline{1S}} = 1/4 (m_{D_{s}}+3m_{D_{s}^{\ast}})$. The 
lattice value of the $D_{s1}(2460)$ bound state position in the infinite volume 
limit ($V\to\infty$) is obtained by an analytical continuation of the scattering 
amplitude combined with L\"uscher's finite volume method. The mass of 
$D_{s1}(2536)$ meson is given directly from the lattice computations without 
resorting to the L\"uscher method. In Fig.~\ref{fig:Ds24602536}, the dashed 
lines represent the threshold for $D^{\ast}K$ in the different approaches 
and the dotted lines are the thresholds for $D^{\ast0}K^{+}$ and 
$D^{\ast+}K^{0}$ in experiment.

The naive quark model predicts that the states corresponding to the 
$D_{s1}(2460)$ and $D_{s1}(2536)$ mesons are almost degenerated, with masses 
close to the experimentally observed mass of the $D_{s1}(2536)$. The inclusion 
of the $1$-loop corrections to the OGE potential does not improve the 
situation, making the splitting between the two states even smaller. Following 
lattice criteria, we couple first the $D^{\ast}K$ threshold in an $S$-wave with 
the two $1^{+}$ $c\bar{s}$ states. One can see in Fig.~\ref{fig:Ds24602536} 
that the state associated with the $D_{s1}(2460)$ meson goes down in the 
spectrum and it is located below $D^{\ast}K$ threshold with a mass compatible 
with the experimental value. The state associated with the $D_{s1}(2536)$ meson 
is almost insensitive to this coupling because it is the $J^{P}=1^{+}$ member 
of the $j_{q}=3/2$ doublet predicted by HQS and thus it couples mostly in a 
$D$-wave to the $D^{\ast}K$ threshold. Lattice QCD has not yet computed the 
coupling in $D$-wave of the $D^{\ast}K$ threshold with the $1^{+}$ $c\bar{s}$ 
sector. This coupling is trivially implemented in our approach. The state 
associated with the $D_{s1}(2460)$ meson experience a very small modification 
because it is almost the $|1/2,1^{+} \!\!\left.\right\rangle$ eigenstate of 
HQS, whereas the state associated with $D_{s1}(2536)$ meson suffers a moderate 
mass-shift approaching to the experimental value. 

Some comments related with the lattice results are due here. The lowest level 
in both lattice ensembles is associated with the physical state 
$D_{s1}(2460)$. This state is below the $D^{\ast}K$ threshold in both lattice 
configurations (Ensemble~(1) and Ensemble~(2)) and it is seen already when 
using only $q\bar{q}$ interpolators. This observation should have important 
consequences in the interpretation of its compositeness. The level is 
down-shifted by about $20\,{\rm MeV}$ (Ensemble (1)) or $30\,{\rm MeV}$ 
(Ensemble (2)) if $D^{\ast}K$ interpolators are included~\cite{Lang:2014yfa}. 
The second state in both ensembles is identified with the $D_{s1}(2536)$ meson. 
In Ensemble~(1), with the heavier pion, the state lies below the $D^{\ast}K$ 
threshold, in strong disagreement with experimental observations. However, in 
the Ensemble~(2), the $D_{s1}(2536)$ state appears above the $D^{\ast}K$ 
threshold. For the two lattice configurations, the effect of coupling the 
$D^{\ast}K$ threshold to both naive $q\bar{q}$ states seems to be small (see 
Fig.~\ref{fig:Ds24602536}). This is expected for the state associated to the 
$D_{s1}(2536)$ but not for the state associated with the $D_{s1}(2460)$ because 
the $D^{\ast}K$ threshold is coupled only in $S$-wave. It is also found in 
lattice computations that the $D_{s1}(2536)$ state is not seen if only 
$D^{\ast}K$ interpolators are used.

Table~\ref{tab:Ds2460P} shows the probabilities of the different Fock 
components in the physical $D_{s1}(2460)$ and $D_{s1}(2536)$ states. When the 
$D^{\ast}K$ threshold is coupled, the meson-meson component is around $50\%$ 
for both $D_{s1}(2460)$ and $D_{s1}(2536)$ mesons. This is in agreement with 
the fact that lattice calculations~\cite{Lang:2014yfa} find similar overlaps of 
the physical states with the quark-antiquark and meson-meson interpolators. 
Moreover, our prediction in this case is in agreement, within errors, with the 
one reported in Ref.~\cite{Torres:2014vna}: the $D_{s1}(2460)$ wave function 
has a probability of $(57\pm21\pm6)\%$ for the $S$-wave $D^{\ast}K$ component.

It is also relevant to realize that the quark-antiquark component in the wave 
function of the $D_{s1}(2536)$ meson holds quite well the $^{1}P_{1}$ and 
$^{3}P_{1}$ composition predicted by HQS. As pointed out in 
Ref.~\cite{Segovia:2009zz}, this is crucial in order to have a very narrow 
state and describe well its decay properties. In Table~\ref{tab:ratios} we 
compare the results obtained in the present calculation with the updated ones 
of Ref.~\cite{Segovia:2009zz}\footnote{The updated results of 
Ref.~\cite{Segovia:2009zz} quoted herein are slightly different from those of 
Ref.~\cite{Segovia:2009zz} since in this work we use the scale-dependent 
strength $\gamma$ of the $^{3}P_{0}$ model obtained in 
Ref.~\cite{Segovia:2012cd}, which is very close but not the same to the value 
used in Ref.~\cite{Segovia:2009zz}.} and the experimental results. 
The theoretical ratios, which pose very demanding constraints to the 
$D_{s1}(2536)$ wave function, are compatible with experiment indicating that 
our mixture for the $D_{s1}(2536)$ wave function describe reasonably well the 
phenomenology of this state. Furthermore, the sophisticated coupled-channel 
study presented herein supports the more phenomenological one performed in 
Ref.~\cite{Segovia:2009zz}.

\begin{table}
\caption{\label{tab:ratios} The total decay width, $\Gamma$, and the branching 
ratios $R_{1}$, $R_{2}$ and $R_{3}$ defined in Ref.~\cite{Segovia:2009zz} for 
the present calculation. We compare with the updated results of 
Ref.~\cite{Segovia:2009zz} (see text for details) and the experimental values 
reported by RPP~\cite{Agashe:2014kda}.}
\begin{ruledtabular}
\begin{tabular}{lccc}
& This work & Updated Ref.~\cite{Segovia:2009zz} & 
Experiment~\cite{Agashe:2014kda} \\
\hline
$\Gamma$ (MeV) & $0.56$ & $0.99$ & $0.92\pm0.03\pm0.04$ \\
$R_1$          & $1.15$ & $1.31$ & $1.18\pm0.16$ \\
$R_2$          & $0.52$ & $0.66$ & $0.72\pm0.05\pm0.01$ \\
$R_3(\%)$      & $14.5$ & $14.1$ & $3.27\pm 0.18\pm 0.37$ \\
\end{tabular}
\end{ruledtabular}
\end{table}

%%%%%%%%%%%%%%%%%%%%%%%%%%%%%%%%%%%%%%%%%%%%%%%%%%%%%%%%%%%%%%%%%%%%%%%%%%%%%%%

\begin{table}[!t]
\caption {\label{tab:lowerDs} Open-flavour strong decay widths, in MeV, and 
branching fractions, in $\%$, of the $D_{s2}^{\ast}(2573)$ meson. Experimental 
data are taken from Ref.~\cite{Agashe:2014kda}.}
\begin{ruledtabular}
\begin{tabular}{lccc}
Channel & $\Gamma_{^{3}P_{0}}$ & ${\cal B}_{^{3}P_{0}}$ & $\Gamma_{\rm exp.}$ \\
                 & (MeV)   & (\%)    & (MeV) \\
\hline
$D^{+}K^{0}$     & $8.02$  & $42.95$ & - \\
$D^{0}K^{+}$     & $8.69$  & $46.54$ & - \\
$D^{\ast+}K^{0}$ & $0.82$  & $4.40$  & - \\
$D^{\ast0}K^{+}$ & $1.06$  & $5.67$  & - \\
$D_{s}^{+}\eta$  & $0.08$  & $0.44$  & - \\
total            & $18.67$ & $100$   & $17\pm4$ \\
\end{tabular}
\end{ruledtabular}
\end{table}

\subsection{The dressed $D_{s2}^{\ast}(2573)$ meson}
\label{subsec:2p}

The $D_{s2}^{\ast}(2573)$ mass and total decay width are predicted well using 
naive quark models and thus this state is commonly expected to be a 
conventional charmed-strange meson. Moreover, the nearest $DK$-type thresholds 
are far enough in order to assume that they do not play an important role in 
the dynamical composition of the $D_{s2}^{\ast}(2573)$ meson.

The same reasoning was followed by the lattice group~\cite{Lang:2014yfa} and 
only quark-antiquark operators in the configuration basis were used in the 
study of the $D_{s2}^{\ast}(2573)$. They also obtain a mass in qualitative 
agreement with experiment confirming that this state can be described well 
within the $c\bar{s}$ picture.

Our predicted mass is shown in Table~\ref{tab:1loopDmesons}, one can see our 
results taking into account the one-gluon exchange potential $(\alpha_{s})$ and 
including its one-loop corrections $(\alpha_{s}^{2})$. In both cases our values 
are slightly higher than experiment but compatible.

We give in Table~\ref{tab:lowerDs} the partial and total strong decay widths of 
the $D_{s2}^{\ast}(2573)$ meson. We show the absolute values in MeV and the 
branching fractions in \%. One can see that the total decay width reported by 
PDG~\cite{Agashe:2014kda} is in excellent agreement with our result. The $DK$ 
channel is clearly dominant with respect the other two possible decay channels, 
$D^{\ast}K$ and $D_{s}\eta$. Therefore, in a coupled-channel calculation the 
mass-shift of the $J^{P}=2^{+}$ ground state would be an effect mainly driven 
by its coupling with the $DK$ threshold. However, in order to do this, the $D$ 
and $K$ mesons should be in a relative $D$-wave and thus carrying extra 
momentum which would imply a small shift.

\begin{figure*}[!t]
\begin{center}
\epsfig{figure=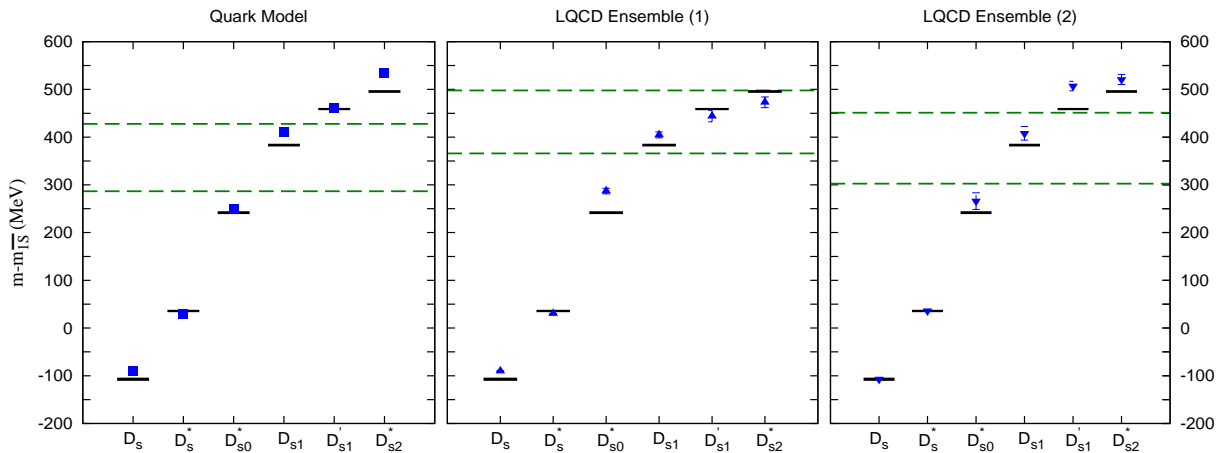,height=0.25\textheight,width=0.90\textwidth}
\caption{\label{fig:Dstotal} Resulting $D_{s}$ spectrum for all channels 
from CQM (squares), LQCD-Ensemble~(1) (up-triangles) and LQCD-Ensemble~(2) 
(down-triangles). The masses are presented with respect to the spin-averaged 
ground state mass $m_{\overline{1S}} = 1/4 (m_{D_{s}} + 3m_{D_{s}^{\ast}})$. 
The black solid lines correspond to experiment. The green dashed lines 
correspond to $DK$ and $D^{\ast}K$ thresholds in the respective approaches. 
Remind that the lattice values of the $D_{s0}^{\ast}(2317)$ and $D_{s1}(2460)$ 
bound state positions in the infinite volume limit ($V\to\infty$) are obtained 
by an analytical continuation of the scattering amplitude combined with 
L\"uscher's finite volume method.}
\end{center}
\end{figure*}

Finally, we show in Fig.~\ref{fig:Dstotal} the low-lying energy spectrum of the 
charmed-strange meson sector and compare with those predicted by the two 
ensembles of lattice QCD. Leaving apart the spectrum obtained using 
Ensemble~(1) which has some features in strong disagreement with experimental 
observations, the spectrum predicted by our coupled-channel quark model and the 
Ensemble~(2) of lattice are in a global agreement and compare quite nice with 
the experimental situation.

%%%%%%%%%%%%%%%%%%%%%%%%%%%%%%%%%%%%%%%%%%%%%%%%%%%%%%%%%%%%%%%%%%%%%%%%%%%%%%%
%%%%%%%%%%%%%%%%%%%%%%%%%%%%%%%%%%%%%%%%%%%%%%%%%%%%%%%%%%%%%%%%%%%%%%%%%%%%%%%

\section{EPILOGUE}
\label{sec:epilogue}

We have performed a coupled-channel computation taking into account the 
$D_{s0}^{\ast}(2317)$, $D_{s1}(2460)$ and $D_{s1}(2536)$ mesons and the $DK$ 
and $D^{\ast}K$ thresholds within the framework of a constituent quark model 
(CQM) whose parameters are largely constraint by hadron observables, from the 
light to the heavy quark sectors.

Our study has been motivated by the fact that recent lattice QCD computations 
need to incorporate explicitly meson-meson operators in their interpolator 
basis in order to obtain correct states for the physical $D_{s0}^{\ast}(2317)$ 
and $D_{s1}(2460)$ mesons. Our method allows to introduce the coupling with the 
$D$-wave $D^{\ast}K$ channel and the computation of the probabilities 
associated with the different Fock components of the physical state, features  
which cannot be addressed nowadays by lattice calculations.

The $D_{s0}^{\ast}(2317)$ meson benefits most from the coupling of the $DK$ 
threshold. The level assigned to it is much higher than the experimental value 
in the naive quark model. However, the $1$-loop corrections to the OGE 
potential brings down this level and locates it slightly above the $DK$ 
threshold. This makes the coupling with the nearby threshold to acquire an 
important dynamical role. When coupling, the level is down-shifted again 
towards the experimental mass of the $D_{s0}^{\ast}(2317)$ meson which is below 
the $DK$ threshold. We predict a probability of $34\%$ for the $DK$ component 
of the $D_{s0}^{\ast}(2317)$ wave function.

The naive quark model predicts that the states corresponding to the 
$D_{s1}(2460)$ and $D_{s1}(2536)$ are almost degenerated with masses 
close to the $D_{s1}(2536)$ mass observed experimentally. The inclusion of the 
$1$-loop corrections to the OGE potential does not improve the situation making 
the splitting between the two states even smaller. When the coupling with $S$- 
and $D$-wave $D^{\ast}K$ threshold is performed, the states associated with the 
physical $D_{s1}(2460)$ and $D_{s1}(2536)$ mesons are in reasonable 
agreement with the experimental situation and lattice findings. We observe that 
the meson-meson component is around $50\%$ for both $D_{s1}(2460)$ and 
$D_{s1}(2536)$ mesons. The $D_{s1}(2536)$ meson appears as the $|3/2,1^{+} 
\!\!\left.\right\rangle$ eigenstate of HQS which is crucial to 
describe its decay properties.

The mass and total decay width of the $D_{s2}^{\ast}(2573)$ meson are predicted 
reasonably well within our quark model approach taking into account only 
quark-antiquark degrees of freedom. We have calculated the partial decay widths 
of this state into open-flavoured mesons. The $DK$ channel is clearly dominant 
with respect the other two possible decay channels, $D^{\ast}K$ and 
$D_{s}\eta$. Therefore, in a coupled-channel calculation the mass-shift of the 
$J^{P}=2^{+}$ ground state would be an effect mainly driven by its coupling 
with the $DK$ threshold. However, in order to do this, the $D$ and $K$ mesons 
should be in a relative $D$-wave and thus carrying extra momentum which would 
imply a small shift.

Finally, our spectrum of the low-lying charmed-strange mesons compares nicely 
with the most updated lattice QCD computation and with the experimental 
situation.

%\appendix
%\section{}

%%%%%%%%%%%%%%%%%%%%%%%%%%%%%%%%%%%%%%%%%%%%%%%%%%%%%%%%%%%%%%%%%%%%%%%%%%%%%%%

\begin{acknowledgments}
We would like to thank M. Albaladejo, J. Nieves and D. Mohler for insightful 
comments. J.S. would like also to express his gratitude to N. Brambilla and A. 
Vairo for a careful reading of the manuscript.
This work has been partially funded by Ministerio de Ciencia y Tecnolog\'\i a 
under Contract no. FPA2013-47443-C2-2-P and by the Spanish Excellence Network 
on Hadronic Physics FIS2014-57026-REDT. P.G.O. acknowledges the financial 
support from the Spanish Ministerio de Economia y Competitividad and European 
FEDER funds under the contract no. FIS2014-51948-C2-1-P. J.S. acknowledges the 
financial support from Alexander von Humboldt Foundation.
\end{acknowledgments}

%%%%%%%%%%%%%%%%%%%%%%%%%%%%%%%%%%%%%%%%%%%%%%%%%%%%%%%%%%%%%%%%%%%%%%%%%%%%%%%

\bibliography{MolecularComponentsDs0Ds1}

\end{document}